\documentclass[a4paper,11pt]{article}
\usepackage{amssymb,amsfonts}
\usepackage{latexsym}
\newtheorem{thm}{Theorem}
\newtheorem{lem}{Lemma}

\newtheorem{prop}{Proposition}
\newcommand{\cmp}{Commun.Math.Phys. $\;$}
\newcommand{\R}{\mathbb{R}}
\newcommand{\C}{\mathbb{C}}
\newcommand{\3}{\ss}
\newcommand{\A}{{\mathfrak A}}
\newcommand{\F}{{\mathfrak F}}
\newcommand{\DK}{{\cal O}}
\newcommand{\AO}{{\mathfrak A}({\cal O})}
\newcommand{\FO}{{\mathfrak F}({\cal O})}
\newcommand{\HR}{{\cal H}}
\newcommand{\K}{{\cal K}}
\newcommand{\G}{\Gamma}
\newcommand{\bewende}{\hspace*{\fill}\nolinebreak\hspace*{\fill}$\Box$
\par\vspace{3mm}}
\title{Braid group statistics in two-dimensional quantum field theory}
\author{C. Adler
\thanks{e-mail: adler@x4u2.desy.de}
\bigskip \\
II. Institut f\"ur Theoretische Physik \\
Universit\"at Hamburg \\
Luruper Chaussee 149 \smallskip \\
D -- 22761 Hamburg}
\date{December 1995}
\begin{document}
\maketitle
\thispagestyle{empty}
\begin{abstract}
Within the framework of algebraic quantum field theory, we construct
explicitly localized morphisms of a Haag-Kastler net in
$1+1$-dimensional Minkowski space showing abelian braid group statistics.
Moreover,
we investigate the scattering theory of the corresponding quantum fields.
\end{abstract}
\section*{Introduction}
The notion of statistics has been of great importance since the early days
of quantum physics, e.g. in connection with Pauli's exclusion principle
or Bose condensation. Originally, one believed permutation group statistics
to be the only possibility to appear; nowadays, one knows that this
is true only in $d \geq 4$ space-time dimensions, while in low dimensional
theories braid group statistics can occur \cite{BF}, \cite{FRS1}.

Here we consider charges localized in some double cone $\DK$ in $1+1$-
dimensional space-time, and the appearance of braid group statistics is
due to the fact that the causal complement of $\DK$ is disconnected, see
e.g. \cite{Haa}.

One of the basic achievements of the theory of superselection sectors
is to give an intrinsic definition of statistics. We use the results
of the abstract analysis to investigate a concrete model showing
abelian braid group statistics.

In section $1$ we outline the algebraic approach to quantum field theory
as initiated by Haag and Kastler; in particular, we define the
statistics operator --- which will play a crucial role in the subsequent
treatment --- and show how this object gives rise to a representation
of the braid group.

After that, in section $2$, we introduce a class of automorphisms
of the algebra of canonical anticommutation relations, namely
the group of Bogoliubov transformations implementable in some Fock
representation, which is an essential tool in the construction of
localized morphisms; in addition, we specify a provisional net of
observables, which does not yet meet all necessary conditions
for the general theory of superselection sectors to be applicable.

We define localized automorphisms of this net in section $3$, using
Bogoliubov operators studied before by Ruijsenaars in connection with
certain integrable field theories \cite{Rui1}.
Moreover, we compute the statistics
and obtain a one-dimensional representation of the braid group.

In section $4$ we extend our provisional net to get one fulfilling
the requirements of superselection theory; furthermore, we extend the morphisms
constructed in the previous section to the latter and see how this gives
rise to an additional quantum number purely topological in nature.

Finally, in section $5$, we discuss the Haag-Ruelle scattering theory
for the corresponding quantum fields, obtaining a close relation between
scalar products of scattering states and the statistics operator.

\section{The algebraic approach}
Besides the Wightman framework \cite{SWig}, algebraic quantum field theory,
which has its origin in the work of Haag, Kastler and Araki \cite{Ar1},
\cite{HK},
has proven successful for the investigation of conceptional questions
in quantum physics.
Since we take the algebraic point of view in this article, we sketch
the main features of this approach; details can be found in the book
of Haag \cite{Haa} or in \cite{Kas} as well as in the original papers
\cite{DHR12}, \cite{DHR34}, \cite{DR1}.

Algebraic quantum field theory is based on the assumption that
a quantum field theory is fixed once the net of observables is specified;
more precisely, denote by ${\cal K}$ the set of open double cones
\footnote{Recall that a double cone is a non-void intersection of
a forward and a backward light cone.} in
Minkowski space; to each double cone $\DK \in {\cal K}$ one associates
a von Neumann algebra $\AO$ in a Hilbert space $\HR$, whose self-adjoint
elements are interpreted as the observables measurable within $\DK$, such
that the following properties are fulfilled (Haag-Kastler axioms):

\begin{description}
\item[1.] {\bf Isotony}:
\begin{eqnarray}
\DK_1 \subset \DK_2 \Longrightarrow \A (\DK_1) \subset \A (\DK_2) \; ,
\end{eqnarray}
i.e. by enlarging the space-time region the algebra of observables
cannot become smaller. Since ${\cal K}$ is directed, one can
consider the algebra
\begin{eqnarray}
\A_{{\rm loc}} := \bigcup_{\DK \in {\cal K}} \AO
\end{eqnarray}
of local observables, which possesses a unique $C^*$-norm $\| . \|_{C^*}$;
by completion with respect to this norm one obtains the so-called
{\it quasilocal algebra}
\begin{eqnarray}
\A := \overline{\A_{{\rm loc}}}^{\| . \|_{C^*}} \; .
\end{eqnarray}
Put differently: $\A$ is the $C^*$-inductive limit of the algebras $\AO$.
\item[2.] {\bf Locality}: The observation that events in space-like
separated regions cannot influence each other is taken into account
by demanding commutativity for algebras associated with space-like separated
double cones ($\times$ denotes space-like separation):
\begin{eqnarray}
\DK_1 \times \DK_2 \Longrightarrow [\A (\DK_1), \A (\DK_2)] = 0
\end{eqnarray}
or equivalently
\begin{eqnarray}
\DK_1 \subset \DK_2' \Longrightarrow \A (\DK_1) \subset \A (\DK_2)' \; .
\end{eqnarray}
Here $\AO '$ means the commutant of $\AO$ and $\DK '$ denotes the causal
complement of $\DK$. Since this axiom mirrors the fact that no signal
can propagate faster than light it is sometimes called
{\it Einstein causality}.
\item[3.] {\bf Haag duality}: Locality can be strengthened by claiming
$\AO$ to be the maximal algebra fulfilling (5), i.e.
\begin{eqnarray}
\AO = \A (\DK ')' \; ,
\end{eqnarray}
where $\A (\DK ')$ denotes the $C^*$-subalgebra of $\A$ generated by
all $\A (\DK_1)$ with $\DK_1 \; \times \; \DK$.
This condition is not always met; in particular, it is violated for the net
of observables of the free Dirac field in two space-time dimensions,
see e.g. \cite{Haa}.
\item[4.] {\bf Covariance}: There exists a representation $\alpha$ of the
Poincar\'e group ${\cal P}_+^{\uparrow}$ by automorphisms of $\A$
such that
\begin{eqnarray}
\alpha_{x,\Lambda} (\AO) = \A ((x,\Lambda) \DK) \; .
\end{eqnarray}
\item[5.] {\bf Spectrum condition}: There is a strongly continuous
unitary representation $(x,\Lambda) \mapsto U(x,\Lambda)$ of
${\cal P}_+^{\uparrow}$ on the
physical Hilbert space $\HR$ implementing $\alpha_{(x,\Lambda)}$, i.e.
\begin{eqnarray}
\alpha_{(x,\Lambda)} (A) = U(x,\Lambda) A U(x,\Lambda)^* \; ,
\end{eqnarray}
such that the generators of $U$ have spectrum in $\overline{V_+}$
(the closure of the forward light cone)
\footnote{This condition is also called ``positivity of energy".};
moreover, there exists a nontrivial vector $\Omega \in \HR$ (unique up to
a phase) --- called the vacuum vector --- which is left invariant by
$U(x,\Lambda)$:
\begin{eqnarray}
U(x,\Lambda) \Omega = \Omega \; .
\end{eqnarray}
\item[6.] {\bf Additivity}: For a covering $\DK_{\lambda}$
of $\DK \in {\cal K}$,
$\AO$ is contained in the von Neumann algebra generated by
$\{ \A (\DK_{\lambda}) \}$:
\begin{eqnarray}
\DK \subset \DK_{\lambda} \Longrightarrow
\AO \subset \bigvee \A (\DK_{\lambda}) \; .
\end{eqnarray}
\end{description}

Besides the net $\DK \mapsto \AO$
\footnote{Note that the local algebras are in general hyperfinite factors
of type $III_1$ --- hence identical up to isomorphism; therefore,
all physical information is contained in the {\it attachment} of the algebras
$\AO$ to the double cones!}
one has to specify the states viewed to be physically realizable or,
equivalently, the representations obtained by GNS construction.
Originally, Borchers investigated all positive energy representations,
but it soon turned out that insurmountable difficulties arise from long range
forces and infrared clouds in case the mass spectrum has no gap.
Doplicher, Haag and Roberts therefore proposed the following criterion:

\vspace{2mm}
{\bf DHR criterion}: {\it Consider those representations $\pi$ which are
unitarily equivalent to the vacuum representation $\pi_0$ when restricted
to the causal complement of a sufficiently large double cone, i.e.
\begin{eqnarray}
\pi|_{\A(\DK')} \cong \pi_0|_{\A(\DK')}
\end{eqnarray}
for sufficiently large $\DK \in {\cal K}$.}

\vspace{2mm}
{\footnotesize {\it Remark}: Unfortunately,\,  ``topological \, charges" \,
(e.g. \, gauge \, charges) \,
are \, ruled \, out \, by \, this criterion
\footnote{Think of electric charge: By Gau\3' law the charge inside an
arbitrarily large sphere can be determined by virtue of a flux
measurement.}.
In case of theories with mass gap the situation has been improved
by the analysis of Buchholz and Fredenhagen \cite{BF}, who considered states
localized in space-like cones. \par}

\vspace{2mm}
Nevertheless, we confine to states satisfying the DHR criterion
in the present paper. In this case there is a unitary
$V:\HR_{\pi} \longrightarrow \HR_0$ such that
\begin{eqnarray}
\pi(A) = V \pi_0 (A) V^* \; , \; \; \; \; \; A \in \A (\DK') \; .
\end{eqnarray}
In what follows we usually identify the quasilocal algebra $\A$ with its
realization in the vacuum Hilbert space, i.e. $A = \pi_0(A)$.
Putting
\begin{eqnarray}
\varrho(A) := V^* A V \; , \; \; \; \; \; A \in \A \; ,
\end{eqnarray}
one realizes that the representation $\pi$ describes a localized
charge exactly if it is unitarily equivalent to a representation
$\pi_0 \circ \varrho$, where $\varrho$ is an {\it endomorphism of $\A$
localized in $\DK$}, i.e.
\begin{eqnarray}
\varrho (A) = A \; \; \; \; \; \forall A \in \A (\DK ') \; .
\end{eqnarray}
The crucial point is the possibility of composing endomorphisms; this way
Doplicher, Haag and Roberts succeeded in analyzing the superselection
structure
\footnote{The physical Hilbert space $\HR$ decomposes into a direct sum
of subspaces under the action of $\A$; these (or rather the associated
equivalence classes of unitary irreducible representations of $\A$) are
called superselection sectors.}.

Two localized morphisms $\varrho_1$, $\varrho_2$
are called {\it equivalent} if the representations
$\varrho_1(\A)$ and $\varrho_2(\A)$ are unitarily equivalent;
mutatis mutandis one defines irreducibility.
A morphism $\varrho$ localized in $\DK$ is {\it transportable} if to each
region $\tilde{\DK}$ obtained from $\DK$ by virtue of a
Poincar\'e transformation
there exists a morphism $\tilde{\varrho}$ localized in $\tilde{\DK}$
equivalent to $\varrho$. Finally, a unitary $U \in \A_{{\rm loc}}$ induces a
localized, transportable automorphism of $\A$ by virtue of
$\sigma_U(A) := UAU^*$ ---
the so-called {\it inner automorphisms}.

By definition, the observables map each sector into itself; in case of
several superselection sectors one is therefore interested in constructing
additional {\it field operators} being local relative to the observables
and inducing transitions between the different sectors.
Following \cite{DHR12},
the {\it field algebra} ${\mathfrak F}$ is obtained from $\A$ by adjoining
localized morphisms $\varrho$ of $\A$.

For permutation group statistics, Doplicher and Roberts were able ---
given the quasilocal algebra $\A$ and the superselection structure ---
to recover a field algebra ${\mathfrak F}$ and a
global gauge group ${\mathfrak G}$
such that $\A$ is the ${\mathfrak G}$-invariant part of ${\mathfrak F}$,
$\A = {\mathfrak F}^{{\mathfrak G}}$ \cite{DR2}.
In the low dimensional case ($d \leq 2$, $d$ the spatial dimension) braid
group statistics may occur and the situation has not yet been completely
clarified, but see \cite{MS}, \cite{Reh2}, \cite{Scho}.

According to the general theory of superselection sectors, for each
$\varrho$ localized in a double cone $\DK$ there exists a unitary
$\epsilon_{\varrho} \in \AO$ commuting with $\varrho^2(\A)$ and fulfilling
\begin{eqnarray}
\epsilon_{\varrho} \varrho(\epsilon_{\varrho}) \epsilon_{\varrho} =
\varrho(\epsilon_{\varrho}) \epsilon_{\varrho}
\varrho(\epsilon_{\varrho}) \; .
\end{eqnarray}
This so-called {\it statistics operator} is given by
\begin{eqnarray}
\epsilon_{\varrho} = U^* \varrho(U) \; ,
\end{eqnarray}
where $U$ is a unitary such that
\begin{eqnarray}
\tilde{\varrho}(A) := U \varrho(A) U^* \; , \; \; \; \; \; A \in \A \; ,
\end{eqnarray}
is localized in a double cone $\tilde{\DK} \subset \DK '$. One easily
checks that a unitary representation of the braid group $B_n$
\footnote{$B_n$ is generated by the identity and elements $\sigma_i$,
$i = 1,\dots , n-1$, fulfilling the {\it Artin relations}
\begin{eqnarray*}
\sigma_i \sigma_j & = & \sigma_j \sigma_i \; \; \; |i-j| \geq 2 \; ; \\
\sigma_i \sigma_{i+1} \sigma_i & = &
\sigma_{i+1} \sigma_i \sigma_{i+1} \; .
\end{eqnarray*}
$B_{\infty}$ is the inductive limit of the $B_n$ with respect to the
canonical inclusions.}
is obtained by virtue of \cite{Fre2}, \cite{Reh1}, \cite{RS}
\begin{eqnarray}
\epsilon_{\varrho}^{(n)} (\sigma_i) := \varrho^{i-1} (\epsilon_{\varrho})
\; , \; \; \; \; \; i = 1, \dots, n-1 \; .
\end{eqnarray}
For $\sigma_i^2 = {\bf 1}$, i.e. $\epsilon_{\varrho}^2 = {\bf 1}$, one recovers
permutation group statistics. Obviously, a localized, transportable
morphism $\varrho$ is in general not invertible, but there exists a
{\it left inverse}, compare e.g. \cite{Rob}, i.e. a positive linear map
$\phi_{\varrho}: \A \longrightarrow {\mathfrak B}({\cal H}_0)$ with
\begin{eqnarray}
{\rm (i)} \; \; \;
\phi_{\varrho}(A \varrho(B)) & = & \phi_{\varrho}(A) B \; , \; \; \; \; \;
A, B \in \A \; ; \\
{\rm (ii)} \; \; \;  \phi_{\varrho}({\bf 1}) & = & {\bf 1} \; .
\end{eqnarray}
In case of an irreducible morphism, $\phi_{\varrho}(\epsilon_{\varrho})$
is a multiple of the identity \cite{Haa}:
\begin{eqnarray}
\phi_{\varrho}(\epsilon_{\varrho}) = \lambda_{\varrho} {\bf 1} \; ;
\end{eqnarray}
$\lambda_{\varrho}$ is called the {\it statistical parameter} and may be
written as $\lambda_{\varrho} = \frac{\omega_{\varrho}}{d_{\varrho}}$,
$|\omega_{\varrho}| = 1$, $d_{\varrho} \geq 1$ ($\omega_{\varrho}$ is the
{\it statistical phase} and $d_{\varrho}$ the {\it statistical dimension}
\footnote{$d_{\varrho}$ is related to the index of the inclusion
$\varrho(\AO) \subset \AO$ by means of ${\rm Ind} \varrho =
d_{\varrho}^2$ \cite{Lon}.}).
We will consider infra the case of $\varrho$ being an automorphism;
then $\epsilon_{\varrho}$ itself is a multiple of the identity, the induced
representation of $B_{\infty}$ is one-dimensional ($d_{\varrho} = 1$)
and $\omega_{\varrho}$ is an arbitrary phase factor.

Exotic statistics has already been investigated by Streater and Wilde
\cite{SWil}
and later by Wilczek \cite{Wil}, who introduced the term ``anyon" for
$d_{\varrho} = 1$ while for $d_{\varrho} > 1$ Fredenhagen, Rehren and Schroer
suggested the term ``plekton"
\footnote{The Greek expression for ``braided".}
in \cite{FRS2}.

\section{Specification of fields and observables}
Let $H$ be a complex Hilbert space, ${\cal C}_0(H)$ the Clifford algebra
over $H$ and denote by ${\cal C}(H)$ the $C^*$-norm completion of
${\cal C}_0(H)$. Furthermore, let $\Psi : H \longrightarrow {\cal C}(H)$
be an antilinear injection fulfilling canonical anticommutation relations:
\begin{eqnarray}
[\Psi(f), \Psi(g)]_+ = 0 \; \; \; \; \; , \; \; \; \; \;
[\Psi(f), \Psi(g)^*)]_+ = \langle f, g \rangle \; {\bf 1} \; .
\end{eqnarray}
Next we introduce a family of automorphisms of ${\cal C} (H)$
important for our purposes. If
\begin{eqnarray}
U (H) :=
\{ U \in {\mathfrak B}(H) \; | \; UU^* = U^*U = {\bf 1} \}
\end{eqnarray}
denotes the group of unitary Bogoliubov operators, then every
$U \in U (H)$ induces an
automorphism of ${\cal C} (H)$ by
\begin{eqnarray}
\alpha_U (\Psi(f)) = \Psi(Uf) \; .
\end{eqnarray}
In addition, for $p \geq 1$ we denote by ${\cal J}_p$ the trace ideal
$\{A \in {\mathfrak B}(H) \; | \; {\rm Tr} \; |A|^p < \infty \}$.

For every projection $P$ in $H$ , there is a unique pure quasifree
\footnote{Recall that a state on ${\cal C} (H)$ is called quasifree if its
n-point functions are given by
\begin{eqnarray}
\omega_A(\Psi(f_n) \dots \Psi(f_1) \Psi(g_1)^* \dots \Psi(g_m)^*) =
\delta_{mn} \; {\rm det} \; ((f_i, A g_j)) \; ,
\end{eqnarray}
where $A$ is an operator on $H$ satisfying $0 \leq A \leq {\bf 1}$.}
gauge invariant state $\omega_P$ on ${\cal C} (H)$ with two point function
\begin{eqnarray}
\omega_P (\Psi(f) \Psi(g)^*) = \langle f, P g \rangle \; ,
\end{eqnarray}
i.e. a Fock state;
let $(\HR_P, \pi_P, \Omega_P)$ denote the associated GNS triple. It is well
known \cite{SS} that an automorphism $\alpha_U$, $U \in U (H)$,
is unitarily implementable in a Fock representation $\pi_P$, i.e.
there exists $\G (U) \in U (\HR_P)$ with
$\pi_P (\Psi (Uf)) = \G (U) \pi_P (\Psi (f)) \G (U)^*$, if and only if
$PU({\bf 1} - P)$ and $({\bf 1} - P)UP$ are Hilbert-Schmidt operators.

Let us now specify the fields and observables necessary for our
considerations. Put $H := L^2 (\R, \C^2)$ and choose a polarization
on $H$ according to the positive and negative part of the spectrum of
$D_m$, the Dirac operator of mass $m>0$; denote the corresponding
projections by $P_+$ resp. $P_-$. Moreover, let $\alpha_t$ be the
automorphism generated by
${\rm e}^{{\rm i}tD_m}$ and $T(x)$ the operator on $H$
representing a translation by $x$.
Denoting by
$\rule[0.1in]{0.25in}{0.01in}$
the conjugate Hilbert space, the one particle space is given by
$\HR_1 := P_+ H \oplus \overline{P_- H}$ and the physical Hilbert space
$\HR$ is precisely the antisymmetric Fock space over $\HR_1$:
\begin{eqnarray}
\HR = {\cal F}_a (\HR_1) \; .
\end{eqnarray}
For $a$ denoting the annihilation operator, we thus obtain a positive
energy representation $\pi$ of ${\cal C}(H)$ on $\HR$
by virtue of
\begin{eqnarray}
\pi (\Psi(f)) = a(P_+f) + a^*(\overline{P_-f}) \; ,
\end{eqnarray}
this being the only positive energy representation of ${\cal C}(H)$
with respect to the dynamics given by $(\alpha_t)_{t \in \R}$ in
the massive case \cite{Wei}.
Let
\begin{eqnarray}
U_2 (H) := \{ U \in U (H) \; | \; P_{\pm} U P_{\mp}
\in {\cal J}_2 (H) \}
\end{eqnarray}
be the group of Bogoliubov operators unitarily implementable in $\pi$
and ${\mathfrak u}_2 (H)$ its complex Lie algebra. If $X \in
{\mathfrak u}_2 (H)$ is self-adjoint,
${\rm e}^{{\rm i}tX}$ is a norm-continuous
one-parameter subgroup of $U_2 (H)$ and we choose the phases
in the corresponding implementers such that we get a strongly continuous
one-parameter group, i.e. ${\rm d} \G (X)$ being the self-adjoint generator,
we have
\begin{eqnarray}
\G ({\rm e}^{{\rm i}tX}) = {\rm e}^{{\rm i}t{\rm d} \G (X)} \; \; \; \; \;
\forall X=X^* \in {\mathfrak u}_2 (H) \; .
\end{eqnarray}
The additive constant in ${\rm d} \G (X)$ is fixed by demanding its
vacuum expectation value to vanish
\footnote{An argument that this can be done is given e.g. in \cite{CR}.}.

\vspace{2mm}
{\footnotesize {\it Remark}: The reader will have noticed that our
formalism to quantize relativistic fermions is equivalent to a more
geometric approach discussed in detail in \cite{PS}.
Roughly speaking, those vectors in Fock space corresponding to pure
quasifree states are identified with sections in a complex line bundle
${\bf DET}^*$ (the dual determinant bundle) over a complex Hilbert
manifold ${\bf Gr} (H)$, the infinite dimensional Grassmannian over $H$.
More precisely, for a Hilbert space $H = H_+ \oplus H_-$ with given
polarization one defines the Grassmannian ${\bf Gr} (H)$ to be the set of
closed subspaces $W$ of $H$ such that
\begin{description}
\item[(i)] $P_+: W \longrightarrow H_+$ is a Fredholm operator;
\item[(ii)] $P_-: W \longrightarrow H_-$ is a Hilbert-Schmidt operator.
\end{description}
${\bf Gr}(H)$ is a smooth complex Hilbert manifold modeled
on the space ${\cal J}_2 (H_+, H_-)$ of Hilbert-Schmidt operators
from $H_+$ to $H_-$; its connected components are labeled by the Fredholm
index of $P_+$. The group $GL(H)$ of bounded invertible operators on $H$
does not act on ${\bf Gr}(H)$; therefore, one passes to the so-called
restricted linear group $GL_2(H)$, which is the complexification of the
real Banach Lie group $U_2(H)$. The action of $U_2(H)$
on ${\bf Gr}(H)$ extends to a holomorphic action of a central extension
$U_2(H)^\sim$ on ${\bf DET}^*$, i.e. one obtains a representation of
$U_2(H)^\sim$ on Fock space. \par}

\vspace{2mm}
Based on previous results of Fredenhagen \cite{Fre1} and
Klaus and Scharf \cite{KS}, Carey, Hurst and
O'Brien \cite{CHB} observed that if one considers the splitting
\begin{eqnarray}
\HR = \bigoplus_{q \in \mathbb{Z}} \HR_q
\end{eqnarray}
of the physical Hilbert space into charge sectors,
the implementer
$\G (U)$ of $U \in$ $U_2 (H)$
\footnote{This is a somewhat sloppy manner of speaking, of course;
however, we will not distinguish between $U$ and the automorphism
$\alpha_U$ induced by it.}
maps $\HR_q$ onto $\HR_{q+q(U)}$, where $q(U)$
is  the  Fredholm  index  of $P_- U P_-$
\footnote{Clearly, $P_+ U P_+$ and $P_- U P_-$ are Fredholm operators
for any $U \in U_2 (H)$.}.
In more detail, they found that if $\G (U)$ acts on a state $\Phi$
of $p$ particles and $q$ antiparticles, these numbers are changed to
$p + {\rm dim \; ker} \; P_- U P_-$ and $q + {\rm dim \; ker} \; P_+ U P_+$.

For a double cone $\DK$ with basis $B_{\DK}$ at time $t$ we define
\begin{eqnarray}
\FO := \{ \pi (\alpha_t (\Psi(f))) \; | \; {\rm supp} \; f \in B_{\DK} \} ''
\end{eqnarray}
to be the associated local field algebra; put
$\F := \overline{\bigcup_{\DK \in \K} \FO}^{{\| \; \|}_{C^*}}$.
Since Fermi fields anticommute rather than commute if localized in
space-like separated regions, Haag duality is not the appropriate notion
of duality but has to be replaced by what is called {\it twisted duality} :
For $F \in \F$ one defines the bosonic resp. fermionic part of $F$ by
\begin{eqnarray}
F_{\pm} := \frac{1}{2} (F \pm {\rm Ad} \; \G (-{\bf 1})(F))
\end{eqnarray}
and sets
\begin{eqnarray}
\FO^{\tau} := \{ F_+ + \G (-{\bf 1}) F_- \; | \; F \in \FO \} \; .
\end{eqnarray}
Then, instead of Haag duality, one has \cite{DHR12}
\begin{eqnarray}
\FO^{\tau} = \F (\DK ')' \; .
\end{eqnarray}
The local observable algebras consist of the $U(1)$-invariant elements
of the local field algebras:
\begin{eqnarray}
\AO := \FO \cap \{ \G ({\rm e}^{{\rm i}\gamma}) \; | \; \gamma \in \R \} ' \; .
\end{eqnarray}
Note that $\DK \mapsto \AO$ is just the net of observables of the free
Dirac field in two space-time dimensions. Unfortunately --- as we
mentioned before --- this net does not fulfil Haag duality: Without
loss of generality we may consider the situation at time $t=0$;
if $\DK, \DK_1, \DK_2$ are double cones such that $\DK_1$ and $\DK_2$
lie in different connected components of $\DK '$ and $f_1$ resp. $f_2$
are test functions with support in $\DK_1$ resp. $\DK_2$, then
$\pi (\Psi(f_1))^* \pi (\Psi(f_2))$ is contained in $\AO '$ but not in
$\A (\DK ') ''$.

\section{Construction of automorphisms}
Generalizing the construction of Binnenhei \cite{Bin}, we will
now obtain localized automorphisms showing (abelian) braid group statistics
of the quasilocal algebra $\A$ associated with the net of observables defined
in the preceding section.

Before going into details, let us briefly outline our strategy.
To begin with, $\pi ({\cal C}(H))$ is contained in $\F$ as a weakly dense
subalgebra and $\pi ({\cal C}(H)^{U(1)})$
\footnote{${\cal C}(H)^{U(1)}$ denotes the $U(1)$-invariant part
of ${\cal C}(H)$, of course.}
is weakly dense in $\A$. Therefore, it is natural to study
automorphisms $\alpha$ of ${\cal C}(H)$ which leave ${\cal C}(H)^{U(1)}$
invariant and extend to automorphisms of $\A$; moreover, the extension
of $\alpha |_{{\cal C}(H)^{U(1)}}$ should be localized in some double cone
$\DK$ and transportable. Finally, as the restriction of $\pi$
to ${\cal C}(H)^{U(1)}$ decomposes into a direct sum of mutually
inequivalent irreducible representations (see above),
\begin{equation}
\pi |_{{\cal C}(H)^{U(1)}} = \bigoplus_{q \in \mathbb{Z}} \pi_q \;
\footnote{$\pi_q$ the restriction of $\pi |_{{\cal C}(H)^{U(1)}}$ to $\HR_q$}
\; ,
\end{equation}
we look for automorphisms connecting the vacuum to the charged sectors, i.e.
\begin{eqnarray}
\pi_0 \circ \alpha |_{{\cal C}(H)^{U(1)}} \cong \pi_q
\end{eqnarray}
for some $q \in \mathbb{Z}^*$. It is known \cite{Bin} that all these
conditions are met for automorphisms induced by a certain class of
Bogoliubov operators $U \in U_2 (H)$, and our morphisms
are precisely of this form in the sense that
\begin{eqnarray}
\alpha_U (A) = \G (U) A \G (U)^* \; , \; \; \; \; \;  A \in \A \; .
\end{eqnarray}

To this end, for $\varepsilon > 0$ let $h_{\varepsilon}$ be an odd,
real-valued, smooth function on $\mathbb{R}$ being equal to $1$ for $x \geq
\varepsilon$ and increasing monotonously inside the interval
$(- \varepsilon, \varepsilon)$. Define unitary multiplication
operators on $H$ by
\begin{eqnarray}
(U(n, \lambda)f)(x) :=
\pmatrix{{\rm e}^{{\rm i} \pi (n + \lambda) h_{\varepsilon} (x)}
& 0 \cr 0 & {\rm e}^{{\rm i} \pi \lambda h_{\varepsilon} (x)}} f(x)
\; , \; \; \; \; \; n \in \mathbb{Z} \; , \; \lambda \in \R \; ,
\end{eqnarray}
and
\begin{eqnarray}
(A(\lambda)f)(x) := \pmatrix{\pi \lambda h_{\varepsilon} (x)
& 0 \cr 0 & \pi \lambda h_{\varepsilon} (x)} f(x)
\; , \; \; \; \; \; \lambda \in \R \; ,
\end{eqnarray}
i.e. $U(0,\lambda) = {\rm e}^{{\rm i} A(\lambda)}$.
In what follows we will write just $U$ instead of $U(n, \lambda)$ whenever
no confusion can arise.
\begin{lem}
\label{Implementierbarkeit}
$P_{\pm} U P_{\mp} \in {\cal J}_2 (H) \; \; \; \forall n \in \mathbb{Z},
\forall \lambda \in \R$ and $m>0$, i.e. the automorphism $\alpha_U$ of
${\cal C}(H)$ induced by $U$ is unitarily implementable in the Fock
representation $\pi$ for all integers $n$ and all real $\lambda$
in the massive case.
\end{lem}
{\it Proof}: Straightforward calculation; we refer to \cite{CHB}, \cite{CR},
\cite{Rui2} for detailed proofs of this statement.
\bewende

{\footnotesize {\it Remark}: For $m=0$ the operators $P_{\pm}UP_{\mp}$ are
Hilbert-Schmidt only for $\lambda \in \mathbb{Z}$ \cite{RW},
\cite{Rui2}. As we shall
see below, this case does not yield interesting morphisms in the sense
that braid group statistics cannot occur; hence we confine to the massive
case. \par}
\begin{lem}
\label{Ladung}
$q(U) = n$.
\end{lem}
{\it Proof}: This follows immediately from the index formulae for
generalized Wiener-Hopf operators proven by Ruijsenaars in \cite{Rui3}.
Namely, he has shown that the index of $P_- U(n,0)P_-$ equals the
winding number of ${\rm e}^{{\rm i} \pi n h_\varepsilon (\cdot)}$
\footnote{By convention, we choose the winding number of
\begin{eqnarray*}
x \mapsto \frac{x-i}{x+i}
\end{eqnarray*}
to be positive.}; by continuity
in $\lambda$, $P_- U(0,\lambda) P_-$ lies in the connected component
of the identity and has therefore vanishing index.
\bewende

\begin{prop}
\label{Transportierbarkeit}
$U$ induces a localized, transportable automorphism $\alpha_U$ of $\A$
by means of $\alpha_U (A) = \G (U) A \G (U)^*$.
\end{prop}
{\it Proof}: For simplicity, we may restrict to the situation at
$t=0$. By unitarity of $U$, $\alpha_U$ maps ${\cal C}(H)^{U(1)}$ into itself;
moreover, the extension of $\alpha_U|_{ {\cal C}(H)^{U(1)}}$ to $\A$
causes no problem.

Concerning localizability, an operator $V \in U_2$ certainly
does induce a morphism localized in $\DK$ if for each connected component
$\Delta_i, i=1,2$, of $\DK '$ there exists $\tau_i \in U(1)$ such that
$V f = \tau_i f$ for all $f \in H$ with ${\rm supp} \; f \in \Delta_i$
\footnote{In fact, this condition is both necessary and sufficient
\cite{Bin}.},
and $U$ obviously shares this property.

Now, if $U$ induces a morphism localized in $\DK$, $U_x := T(x)UT(-x)$
generates one localized in $(\DK +x)$. We claim that these morphisms
are equivalent by means of a unitary element of a local observable algebra.
To see this, consider the operator $\G (U_xU^*)$.
Clearly, $\G (U_xU^*)$ maps the charge sectors $\HR_q$ into themselves and
commutes with global gauge transformations $\G ({\rm e}^{{\rm i}\gamma})$.
Let $\tilde{\DK} \supset \DK \cup (\DK + x)$;
then $\G (U_xU^*)$ is contained in
$\A (\tilde{\DK})$: it remains to be shown that $\G (U_xU^*)
\in \F(\tilde{\DK})^{\tau} = \F (\tilde{\DK} ')'$; choosing $f \in H$
such that ${\rm supp} \; f \cap B_{\tilde{\DK}} = \emptyset$, we have
\begin{eqnarray*}
\G (U_xU^*) \pi (\Psi(f)) \G (U_xU^*)^* & = & \pi (\Psi(U_xU^* f)) \\
& = & \pi (\Psi(f)) \; .
\end{eqnarray*}
\bewende

{\footnotesize {\it Remark}: The fact that the charge transfer operators
lie in a local observable algebra is not clear a priori since
this is usually shown using Haag duality --- which is violated here. \par}

\vspace{2mm}
To summarize, we succeeded in constructing localized, transportable
automorphisms of $\A$ connecting the vacuum to the different charged sectors.
Our next aim is to compute the corresponding statistics operator
$\epsilon_{\alpha_U}$. To do this, we need two more lemmata.
\begin{lem}
\label{AllgemeineVertauschungsregeln}
\begin{description}
\item[(i)] Let $X, Y \in {\mathfrak u}_2(H)$ be self-adjoint operators
such that $[X, Y] = 0$. Then
\begin{eqnarray}
\G ({\rm e}^{{\rm i}X}) \G ({\rm e}^{{\rm i}Y}) =
{\rm e}^{- c(X,Y)} \G ({\rm e}^{{\rm i}Y}) \G ({\rm e}^{{\rm i}X}) \; ,
\end{eqnarray}
where $c(X,Y) := {\rm Tr} \; (P_-XP_+YP_- - P_-YP_+XP_-)$.
\item[(ii)] For $V \in U_2(H)$
and $X = X^* \in {\mathfrak u}_2(H)$
with $[V, X] = 0$,
\begin{eqnarray}
\G ({\rm e}^{{\rm i}X}) \G (V) =
{\rm e}^{{\rm i} (\G (V) \Omega, {\rm d} \G(X) \G (V) \Omega)}
\G (V) \G ({\rm e}^{{\rm i}X})
\end{eqnarray}
holds true.
\end{description}
\end{lem}
{\it Proof}: That $\G$ is a projective representation of $U_2(H)$
with (Lie group) cocycle $C(X, Y) =
{\rm e}^{-c(X, Y)}$ is well known, of course
\cite{CR}, \cite{PS}; (ii) follows by uniqueness of the implementers \cite{CR}.
\bewende

\vspace{2mm}
{\footnotesize {\it Remark}: Actually, $c(X, Y)$ is the (Lie algebra)
cocycle associated with ${\rm d} \G$ fixing the central extension and known
to physicists as the {\it Schwinger term} occuring in the current
algebra. \par}

\vspace{2mm}
In addition, denoting by $\DK_x$ the double cone with basis $(x - \varepsilon,
x + \varepsilon)$ at time $t=0$, we obtain
\begin{lem}
\label{SpezielleVertauschungsregeln}
For $\DK_x$ space-like to $\DK_{x '}$, one has
\begin{description}
\item[(i)]
\begin{equation}
\G (U_x(n,0)) \G (U_{x'}(n',0)) = (-1)^{nn'}
\G (U_{x'}(n',0)) \G (U_x(n,0)) \; ;
\end{equation}
\item[(ii)]
\begin{equation}
c(A_x(\lambda), A_{x'}(\lambda ') = 0 \;
\footnote{$A_x := T(x)AT(-x)$} \; ;
\end{equation}
\item[(iii)]
\begin{equation}
(\G (U_x(n,0))\Omega, {\rm d} \G(A_{x'}(0,\lambda))
\G (U_x(n,0))\Omega) = \pi n \lambda {\rm sign} \; (x - x') \; .
\end{equation}
\end{description}
\end{lem}
{\it Proof}: Again, (i) is well known \cite{PS}; to prove (ii), notice
that the relevant convolution kernels are given by
\begin{eqnarray*}
P_{\delta}AP_{-\delta} (\theta, \theta ') = N \delta \hat{h}_{\varepsilon}
[\delta (\sinh{\theta} + \sinh{\theta'})]
\sinh{\left( \frac{\theta + \theta'}{2} \right)}
\; , \; \; \; \; \; \delta = +, - \; ,
\end{eqnarray*}
where $\hat{h}_{\varepsilon}$ is the (distributional) Fourier transformation
of $h_{\varepsilon}$, $\theta$ denotes the rapidity (recall that
the momentum $p$ is related to the rapidity by $p(\theta) = m \sinh{\theta}$)
and $N$ is some constant; since $h_{\varepsilon}$ is an odd function, we see
that $P_+AP_- = P_-AP_+$. (iii) is obvious.
\bewende

Now we are in a position to state
\begin{prop}
\label{Statistik}
$\epsilon_{\alpha_U} =
(-1)^n {\rm e}^{{\rm i} 2 \pi n \lambda {\rm sign} \; (x' - x)}
{\bf 1}$.
\end{prop}
{\it Proof}: By definition, we have
\begin{eqnarray*}
\epsilon_{\alpha_U} =
\G (U_xU_{x'}^*)^* \G (U_x) \G (U_xU_{x'}^*) \G (U_x)^* \; .
\end{eqnarray*}
Writing $\G (U_x)$ as $C(U_x(n,0), U_x(0,\lambda))^{-1} \G (U_x(n,0))
\G (U_x(0,\lambda))$
\footnote{$C(.,.)$ denoting again the (Lie group) cocycle corresponding
to the projective representation $\G$.},
the statement follows immediately from lemmata
\ref{AllgemeineVertauschungsregeln} and
\ref{SpezielleVertauschungsregeln}.
\bewende

\vspace{2mm}
{\footnotesize {\it Remark}: Consider the case of automorphisms
$\alpha_U$ resp. $\alpha_{U'}$ induced by Bogoliubov operators
$U = U(n, \lambda)$ resp. $U' = U(n', \lambda ')$; using the fact that
localized, transportable morphisms commute if localized in space-like
separated regions (see e.g. \cite{Rob}), the statistics operator is again
a multiple of the identity; namely, we obtain
\begin{eqnarray}
\epsilon (\alpha_U, \alpha_{U'}) = (-1)^{nn'}
{\rm e}^{{\rm i} \pi (n \lambda ' + n' \lambda) {\rm sign} \; (x' - x)}
{\bf 1} \; .
\end{eqnarray}
\par}

\section{Extension of $\DK \mapsto \AO$}
In this section we extend the net $\DK \mapsto \AO$ to obtain one fulfilling
Haag duality and investigate the superselection structure of the latter.

\vspace{2mm}
{\footnotesize {\it Remark}: Since $\G (U_x (0, \lambda))$ is contained
in $\A(\DK_x')'$ but not in $\A(\DK_x)$, one way to extend the net of
observables is to adjoin $\G (U_x (0, \lambda))$ to $\A(\DK_x)$ and consider
the von Neumann algebra such generated as the new local observable algebra;
this is equivalent to taking the dual net $\A^d (\DK) := \A (\DK ')'$.
Unfortunately, the operators $\G (U_x (0, \lambda))$ are localized
only in half-spaces (after performing a global gauge transformation),
and one has to study algebras $\A (W_{\pm})$ associated with wedge regions
\begin{eqnarray}
W_{\pm} := \{ x \in \R \; | \; |x^0| < \pm x^1 \} \; .
\end{eqnarray}
For these algebras Haag duality is a consequence of twisted duality
\cite{DHR12}:
\begin{eqnarray}
\A (W_+ + x)' = \A (W_- + x)'' \; .
\end{eqnarray}
In this case the natural extension of $\alpha_U$
may be interpreted as a soliton
\footnote{Note that in our case the different vacua are connected by
an internal symmetry; for a generalization we refer to \cite{Fre4}.}. \par}

\vspace{2mm}
The extension we want to study is analogous to the universal
algebra introduced by Fredenhagen, Rehren and Schroer in the context
of conformal field theory \cite{FRS2}. Namely, we define
\begin{eqnarray}
\A_1 (\DK) & := & \AO \; , \\
\A_1 (\DK ') & := & \AO ' \; .
\end{eqnarray}
By construction, this net
\footnote{To be honest, the notion of a net is used only for directed
index sets; what we have defined is rather a precosheaf. However, we
ignore such subtleties and keep on using the term net.}
fulfils Haag duality. The algebras $\A_1 (\DK)$ and $\A_1 (\DK ')$
generate a $C^*$-algebra $\A_1$ uniquely determined by the following
properties:
\begin{description}
\item[(i)] There exist unital embeddings $\imath^I: \A_1(I) \hookrightarrow
\A_1$ such that for all
$I, J \in \Sigma := \{ \DK, \DK ' \; | \; \DK \in \K \}$
one has
\begin{eqnarray}
\imath^J |_{\A_1(I)} = \imath^I \; , \; \; \; \; \; I \subset J \; ,
\end{eqnarray}
and $\A_1$ is generated by the algebras $\imath (\A_1(I))$.
\item[(ii)] For every family $(\pi^I)_{I \in \Sigma} \; , \;
\pi^I: \A_1(I) \longrightarrow {\mathfrak B} (\HR_{\pi})$, of normal
representations with
\begin{eqnarray}
\pi^J |_{\A_1(I)} = \pi^I \; , \; \; \; \; \; I \subset J \; ,
\end{eqnarray}
there is a unique representation $\pi$ of $\A_1$ in $\HR_{\pi}$ such that
\begin{eqnarray}
\pi \circ \imath^I = \pi^I \; .
\end{eqnarray}
\end{description}
Note that the vacuum representation $\pi_0$ of $\A_1$ induced by
$\imath^I (A) \mapsto A$ is in general not faithful.

Our next task is to extend $\alpha_U$ to $\A_1$. For notational
simplicity, let us consider the situation at $t=0$. Since we already know
how $\alpha_U$ operates on local algebras $\A_1(\DK)$, only its
action on $\A_1(\DK ')$, i.e. on the commutants $\AO '$, remains to be
determined. Let $\alpha_U$ be localized in $\DK$ and let $\DK_x$
\footnote{Note that the restriction to double cones
with basis $(x - \varepsilon, x + \varepsilon)$ does not mean any
loss of generality.}
be another double cone. Then $\alpha_U$ is unitarily equivalent to
a morphism localized in $\DK_x$ by means of
\begin{eqnarray}
{\mathfrak T} := \G (U_x U^*) \; .
\end{eqnarray}
As in the proof of proposition \ref{Transportierbarkeit}, we see that
this charge transfer operator is contained in a local algebra
$\A_1 (\tilde{\DK}) \; , \; \tilde{\DK} \; \supset \; \DK \cup \DK_x$.
We define
\begin{eqnarray}
\alpha_U^{\A_1 (\DK_x ')} :=
{\rm Ad} \; {\mathfrak T}^* |_{\A_1 (\DK_x ')} \; .
\end{eqnarray}
We claim that this is well-defined: For if $A$ is in $\A_1 (\DK_{x_1} ')$
as well as in $\A_1 (\DK_{x_2} ')$, both ${\rm Ad} \; \G (U_{x_1})$ and
${\rm Ad} \; \G (U_{x_2})$ act trivially on $A$; therefore,
\begin{eqnarray}
\alpha_U^{\A_1 (\DK_{x_1} ')} (A) = \alpha_U^{\A_1 (\DK_{x_2} ')} (A) \; .
\end{eqnarray}
\begin{thm}
\label{Erweiterung}
The extensions to $\A_1$ of morphisms $\alpha_U$ resp. $\alpha_{U'}$
induced by Bogoliu\-bov operators $U = U(n, \lambda)$ resp.
$U' = U(n', \lambda ')$ are inequivalent provided that $n \neq n'$ or
$\lambda \neq \lambda ' \; \; ({\rm mod} \; 2 \mathbb{Z})$.
\end{thm}
{\it Proof}: Obviously, $\alpha_U$ and $\alpha_{U'}$ cannot be equivalent
if $n \neq n'$, as in this case they connect the vacuum to different
charge sectors. Moreover, for a net fulfilling Haag duality, the notions
of unitary equivalence and inner equivalence coincide: For there exists
a double cone $\DK$ containing the supports both of
$\alpha_U$ and $\alpha_{U'}$, and the morphisms act trivially on
$\A_1 (\DK ')$; but then the unitary operator providing the equivalence
commutes with $\A_1 (\DK ')$ and therefore is in $\A_1 (\DK)$.
Now, according to the general theory of superselection sectors, if
$\alpha_U$ and $\alpha_{U'}$ were inner equivalent, their statistics
operators had to coincide.
\bewende

Since we did not enlarge the local algebras themselves but only
redefined the algebras associated with the (unbounded) causal complements
of double cones, the additional quantum number $\lambda$ is of a
topological type and might be thought of as a ``charge at infinity".

\section{Scattering theory}
In this final section we assume $n=1$. As we mentioned in section $1$,
the theorem of Doplicher and Roberts, yielding a field algebra $\F$ once
the quasilocal algebra $\A$ and the superselection structure are known,
does not apply to our case; therefore, we have to use a substitute ---
the so-called {\it field bundle} ${\cal F}$ \cite{DHR34} --- to describe
charged fields interpolating between different sectors. In this framework,
state vectors
are pairs $\{ \varrho, \phi \}$, $\phi \in \HR$ and $\varrho$ a localized
morphism of $\A$. Furthermore, fields are pairs $\{ \varrho, A \}$,
$A \in \A$, and act on state vectors by virtue of
\begin{eqnarray}
\{ \varrho, A \} \{ \varrho ', \phi \} =
\{ \varrho ' \varrho, \varrho ' (A) \phi \} \; .
\end{eqnarray}
A field $\{ \varrho, A \}$ is localized in a double cone $\DK$ if
\begin{eqnarray}
AB = \varrho (B) A \; , \; \; \; \; \; B \in \A (\DK ') \; .
\end{eqnarray}
Finally, for any two localized morphisms $\varrho_1$ and $\varrho_2$,
one defines the set $\mathfrak{I} (\varrho_2 | \varrho_1)$ of
intertwiners
from $\varrho_1$ to $\varrho_2$:
\begin{eqnarray}
\mathfrak{I} (\varrho_2 | \varrho_1) := \{ T \in \mathfrak{B} (\HR_0) \; |
\; \varrho_2(A) T = T \varrho_1(A) \; \; \forall A \in \A \} \; .
\end{eqnarray}

\vspace{2mm}
{\footnotesize {\it Remark}: At this point, let us comment on a difficulty
showing up if one considers the universal algebra for conformal field theories
living on the circle, i.e. $\A_1 = \A_1 (S^1)$, namely, the existence
of global self-intertwiners \cite{FRS2}.
The problem is that the statistics operator
may depend on the choice of a ``point at infinity" $\xi \in S^1$. For if
$\varrho$ is localized in an interval $J$, $\sigma$ is localized in an
interval $I$ and $\xi \in J' \cap I'$, the statistics operator
$\epsilon_{\xi}(\varrho, \sigma)$ does not change as long as $\xi$
varies continuously; but $J' \cap I'$ may consist of two connected
components $\Delta_1, \Delta_2$. Choosing $\xi \in \Delta_1$ and $\zeta \in
\Delta_2$, the corresponding statistics operators coincide
only if the monodromy is trivial!

Global self-intertwiners arise as follows: Let $\varrho$ and $\sigma$
be localized in $I$ and $\tilde{\varrho}$ a morphism equivalent to
$\varrho$ localized in $J$; then $\epsilon (\varrho, \sigma)$ and
$\epsilon (\sigma, \varrho)$ coincide for $\xi$ and $\zeta$. Putting
\begin{eqnarray}
\A_{\xi} := \overline{\{ \A (I) \; | \; I \in S^1 \; , \xi \notin I \}} \; ,
\end{eqnarray}
both $\epsilon (\varrho, \sigma)$ and $\epsilon (\sigma, \varrho)$
are in $\A_{\xi} \cap \A_{\zeta}$; by Haag duality, there exists
an intertwiner $S: \pi_0 \varrho \longrightarrow \pi_0 \tilde{\varrho}$
contained in $\pi_0 (\A_{\xi})$ as well as in $\pi_0 (\A_{\zeta})$.
Denoting the corresponding preimages by $S_1 \in \A_{\xi}$ resp.
$S_2 \in \A_{\zeta}$, a global self-intertwiner from $\varrho$ to $\varrho$
is given by
\begin{eqnarray}
S_{\varrho} := S_1^* S_2 \; .
\end{eqnarray}
The crucial point is that $S_{\varrho}$ is trivial in the vacuum
representation while $\pi_0 \sigma (S_{\varrho})$ is the monodromy operator:
\begin{eqnarray}
\pi_0 \sigma (S_{\varrho}) = \pi_0 (\epsilon (\varrho, \sigma)
\epsilon (\sigma, \varrho)) \; .
\end{eqnarray}
Roughly speaking, the choice of a ``point at infinity" determines
whether $I$ lies to the right of $J$ or vice versa. For theories
on the real line this difficulty is absent due to the fact that
a ``point at infinity" is given a priori. \par}

\vspace{2mm}
For convenience of the reader not familiar with our approach, let us
now sketch the Haag-Ruelle programme; details may be found e.g. in
\cite{DHR34}, \cite{FGR}.

Take $B' \in \A_1 (\DK)$ for some double cone $\DK$ and let
${\bf B'} = \{ \varrho, B' \}$ be the corresponding field.
Moreover, let $g$ be a Schwartz function on Minkowski space whose Fourier
transform intersects the spectrum of ${\bf B'} \Omega$ only on the
mass hyperboloid $p^2 = m^2$. Define a single particle state by virtue of
\begin{eqnarray}
{\bf B} := \int {\rm d}^2 x g(x) \alpha_x ({\bf B'}) \; .
\end{eqnarray}
Our choice of $g$ and $B'$ guarantees ${\bf B}$ to be almost local;
if
\begin{eqnarray}
f(x) = \int {\rm d} p \hat{f}(p) {\rm e}^{{\rm i}(px^1 - E_pt)}
\; \; \; \; \; \; \; \; \; \;
(\hat{f} \in {\cal C}_0^{\infty} \; , \; \; E_p = \sqrt{p^2 + m^2})
\end{eqnarray}
is a positive
energy solution of the Klein-Gordon equation for mass $m$, one puts
\begin{eqnarray}
{\bf B}_f (t) := \int_{x^0 = t} {\rm d} x^1 f(x) \alpha_x ({\bf B})
\end{eqnarray}
and obtains an eigenvector of the mass operator $M^2$ with eigenvalue
$m^2$ (i.e. a one particle state vector) by means of
\begin{eqnarray}
\Psi := {\bf B}_f (t) \Omega = \{ \varrho, \psi \} \; ;
\end{eqnarray}
$\Psi$ does not depend on $t$. The next step is to define
multiparticle states by means of
\begin{eqnarray}
{\bf B}_{f_n} (t) \dots {\bf B}_{f_1} (t) \Omega \; ;
\end{eqnarray}
for reasons of convergence one assumes the functions $f_i$ to have
mutually disjoint supports in rapidity space. Then one proves strong
convergence of the expressions
\begin{eqnarray}
\lim_{t \rightarrow - \infty} {\bf B}_{f_n} (t) \dots {\bf B}_{f_1} (t) \Omega
=: \Psi_n \times_{in} \dots \times_{in} \Psi_1
\end{eqnarray}
resp.
\begin{eqnarray}
\lim_{t \rightarrow + \infty} {\bf B}_{f_n} (t) \dots {\bf B}_{f_1} (t) \Omega
=: \Psi_n \times_{out} \dots \times_{out} \Psi_1
\end{eqnarray}
and interpretes them as incoming resp. outgoing scattering states.

For later use we state without proof a version of the well known
cluster theorem \cite{FGR}:
\begin{lem}
\label{Cluster}
Let ${\bf B}_i = \{ \varrho_i, B_i \} \in {\cal F} (\DK_i) \; , \;
i = 1, \dots , 4$, such that $\DK_1 \cup \DK_3 \; \times \;
\DK_2 \cup \DK_4$. Furthermore, let $T$ be an intertwiner from
$\varrho_1 \varrho_2$ to $\varrho_3 \varrho_4$ and define
\begin{eqnarray}
\tau := {\rm sup} \{ |t| \; | \; \DK_1 \cup \DK_3 \; + \; (t, 0) \subset
(\DK_2 \cup \DK_4)' \} \; .
\end{eqnarray}
Assuming $\varrho_4$ to be irreducible with finite statistics (i.e.
$d_{\varrho_4} < \infty$) and (unique) right inverse $\chi_4$
\footnote{The notion of a right inverse is analogous to that of a left
inverse introduced in section $1$.}
and denoting by $\{ W_j \}$ a (possibly empty) orthonormal basis of
the Hilbert space of local intertwiners from $\varrho_4$ to $\varrho_2$,
one has
\begin{eqnarray}
| ({\bf B}_4 {\bf B}_3 \Omega, T {\bf B}_2 {\bf B}_1 \Omega) -
\sum_j ({\bf B}_3 \Omega, \chi_4 (T \varrho_1 (W_j)) {\bf B}_1 \Omega)
({\bf B}_4 \Omega, W_j^* {\bf B}_2 \Omega) | \nonumber \\
\leq {\rm e}^{-\mu \tau}
\prod_i \| {\bf B}_i \| \; .
\; \; \; \; \; \; \; \; \; \; \; \; \; \; \; \; \; \; \; \;
\; \; \; \; \; \; \; \; \; \; \; \; \; \; \; \; \; \; \; \;
\end{eqnarray}
\end{lem}

To apply this programme to our morphisms $\alpha_U$, we have to check
that they induce massive single particle representations; but this is clear:
Covariance is a consequence of covariance of the Bogoliubov operators $U_x$:
\begin{eqnarray}
U_{x+a} = T(a) U_x T(-a) \; ,
\end{eqnarray}
and the generators of translations coincide with those for the free field;
hence there exists an isolated mass shell.

Let $\{ \Psi_i \} = \{ {\bf B}_{f_i} (t) \Omega \} =
\{ \{ \varrho_i, \psi_i \} \} \; , \; i=1, \dots n$, be a set of normed
single particle states with mutually disjoint supports in rapidity space,
where all $\varrho_i$ are taken to be copies of $\alpha_U$. Without loss
of generality we assume the support of $f_i$ in rapidity space left to that
of $f_{i+1}$. Hence for $t \rightarrow - \infty$ we have
\begin{equation}
(\Psi_n, \DK_n) < \dots < (\Psi_1, \DK_1) \;
\footnote{Here $\DK_i$ denotes the ``support" of $\Psi_i$, i.e.
${\bf B}_i \in {\cal F} (\DK_i)$, and $\DK_i < \DK_j$ if $\DK_i$ is left
to $\DK_j$.} \; ,
\end{equation}
while for $t \rightarrow + \infty$ we obtain
\begin{eqnarray}
(\Psi_1, \DK_1) < \dots < (\Psi_n, \DK_n) \; .
\end{eqnarray}
These incoming resp. outgoing configurations are linked by means of the
unitary transformation
\begin{eqnarray}
S(\lambda) = (- {\rm e}^{{\rm i} 2 \pi \lambda})^{\sum_{i<j} 1} {\bf 1} =
(- {\rm e}^{{\rm i} 2 \pi \lambda})^{\frac{n(n-1)}{2}} {\bf 1} \; .
\end{eqnarray}
Applying lemma \ref{Cluster} recursively, we have shown
\begin{prop}
\begin{eqnarray}
((\Psi_n, \DK_n) \times_{out} \dots \times_{out} (\Psi_1, \DK_1),
S(\lambda) (\Psi_n, \DK_n) \times_{out} \dots \times_{out} (\Psi_1, \DK_1))
\nonumber \\
= (- {\rm e}^{{\rm i} 2 \pi \lambda})^{\frac{n(n-1)}{2}} \; .
\; \; \; \; \; \; \; \; \; \; \; \; \; \; \; \; \; \; \; \;
\; \; \; \; \; \; \; \; \; \; \; \; \; \; \; \; \; \; \; \;
\end{eqnarray}
\end{prop}

{\footnotesize {\it Remark}: Slightly generalizing the above consideration,
let us take different values of $\lambda$ in the single particle states
$\Psi_i = \{ \varrho_i, \psi_i \}$,
i.e. $\varrho_i = \alpha_{U(1, \lambda_i)}$, such that $\sum \lambda_i =
\Lambda$ for fixed $\Lambda$. In this case, incoming and outgoing
scattering states are linked by
\begin{eqnarray}
S(\Lambda) = (-1)^{\frac{n(n-1)}{2}} {\rm e}^{{\rm i} \pi (n-1) \Lambda} \; ;
\end{eqnarray}
put differently: The phase factor occuring in scalar products
of scattering states does only depend on the ``total charge" $\Lambda$. \par}

\vspace{2mm}
This result may be interpreted as follows: The morphisms $\alpha_U$
describe free anyons with scattering determined by the statistical
parameter. Following a proposal of Swieca \cite{Schr1} to split
the $S$-matrix into a kinematical and a dynamical part, the dynamical
one is trivial in our case since the phase factor does not contribute
to the scattering cross section.

\subsection*{Acknowledgement}
It is a pleasure to thank Prof. Fredenhagen not only for initiating
this line of research but also for many helpful discussions. Financial support
by the Graduiertenkolleg ``Theoretische Elementarteilchenphysik" is
gratefully acknowledged.

\end{document}